\documentclass[twocolumn,prl,showpacs,amsmath,amssymb,floatfix,superscriptaddress]{revtex4}
\usepackage[latin9]{inputenc}
\usepackage{array}
\usepackage{amstext}
\usepackage{graphicx}
\usepackage{wasysym}
\usepackage{esint}

\makeatletter

\providecommand{\tabularnewline}{\\}

\pdfoutput=1

\usepackage{dcolumn}
\usepackage{bm}

\makeatother

\begin{document}
\title{Pressure-induced high temperature superconductivity in H$_{3}$X (X=As,
Se, Br, Sb, Te and I)}
\author{P.-H. Chang}
\author{S.Silayi}
\author{D.A. Papaconstantopoulos}
\email{pchang8@gmu.edu}

\affiliation{Department of Computational and Data Sciences, George Mason University,
Fairfax, VA}
\author{M.J. Mehl}
\affiliation{US Naval Academy, Annapolis, MD}
\begin{abstract}
The discovery of high critical temperature $T_{c}$ superconductivity
in highly compressed H$_{3}$S has opened up the question of searching
for strong electron-phonon coupling in the hydrides outside the transition
metal series. The specific objective of this work is to explore the
possibility of discovering a material that exceeds the superconducting
transition temperature of H$_{3}$S. Our study includes the materials
H$_{3}$X (X=As, Se, Br, Sb, Te, and I), is limited to the $Im\overline{3}m$
crystal structure. The procedure we adopt involves performing linearized
augmented plane wave (LAPW) calculations for many different volumes
to compute the electronic densities of states and their pressure variation.
This is combined with Quantum-ESPRESSO (QE) calculations from which
we obtain the phonon frequencies and the electron-phonon coupling
constant $\lambda$, and followed by applying the multiple scattering-based
theory of Gaspari and Gyorffy (GG) to obtain the Hopfield parameters
and the McMillan-Allen-Dynes theory. It should be stressed that the
GG approach decouples the electronic contribution to $\lambda$ from
the corresponding phonon contribution, and provides additional insights
for the understanding of superconductivity in these materials. Based
on our analysis, the hydrogen is the main contributor to the $T_{c}$
in these materials as it makes up $75\sim80$ \% of the total $\lambda$.
Our calculations for H$_{3}$Se and H$_{3}$Br give a $T_{c}${\normalsize{}{}{}$\sim100$
K. For the other materials in our study we find that H$_{3}$As is
unstable and H$_{3}$Sb, H$_{3}$Te and H$_{3}$I have small values
of the McMillan-Hopfield paramters which makes it unlikely to give
high $T_{c}$}. However, according to both of our rigid band model
and virtual crystal calculations, we predict a $T_{c}\sim150$ K for
H$_{3}$Br with a small amount of hydrogen doping. Our basic conclusion
is that the materials studied here could not reach very high $T_{c}$
because the Hopfield parameters, which are the strongest contributor
to high $T_{c}$, are not large enough. 
\end{abstract}
\maketitle

\section{Introduction}

Recently high temperature superconductivity at temperatures exceeding
200 K was predicted by Duan {\em et al.} \cite{Duan2014} at extreme
pressures above 200 GPa in H$_{3}$S in the $Im\overline{3}m$ crystal
structure.\cite{R3M} The prediction was immediately confirmed experimentally
by Drozdov {\em et al.} \cite{Drozdov2015}. This breakthrough
has motivated numerous theoretical and experimental studies \cite{Fan2016,Flores-Livas2016,Ge2016,Liu2017,Papaconstantopoulos2017,Papaconstantopoulos2017a,Zhang2015b,Zhong2016,Heil2015,2016NatSR...622873M,2018arXiv181008191M,2018arXiv181201561D,Bernstein2015,Nicol2015,Akashi2015,Papaconstantopoulos2015}
and the consensus developed that conventional BCS electron-phonon
coupling is in play. Researchers in this field are exploring other
elements to stabilize hydrogen at high pressures and already have
been reports of near room temperature (RT) superconductivity in the
compound H$_{10}$La \cite{2018arXiv181201561D,Liu2017}. The idea
of metallization of hydrogen that was proposed long ago by Wigner
and Huntington \cite{Wigner1935} has been pursued vigorously and
Ashcroft's prediction \cite{Ashcroft1968} of RT superconductivity
in metallic hydrogen under high pressures is getting close to reality.

Using the Gaspari--Gyorffy (GG) theory \cite{Gyorffy1972}, which
is the basis of the present work, Papaconstantopoulos and Klein \cite{PAPA1977}
predicted the electron-phonon coupling $\lambda=1.86$ and superconducting
temperature $T_{c}=234$ K at a pressure of $460$ GPa for metallic
hydrogen.

However, metallizing hydrogen requires extremely high pressure \cite{Wigner1935}.
Recent theoretical studies also suggest that it would require pressure
at roughly 500 GPa \cite{McMahon2012}. The hydrides are thus introduced
as an alternative which offer a rather satisfactory trade-off since
they could form metallic states at much lower pressure.

The hydrides are considered as unusual but conventional superconductors
since their behavior can be explained with traditional electron-phonon
interaction while a few details differ from the conventional ones
\cite{GorKov2018}. A comprehensive review of superconductivity in
hydrides is given by Zurek et al \cite{Bi2018}. H$_{3}$S is a prominent
example because of its optimal electronic states and the separation
of the acoustic from the optical phonon modes. It is believed that
sulfur lacks a specific role in terms of its contribution to enhancing
superconductivity but it helps hydrogen forming metallic states. To
carry this idea forward, there have been several attempts at targeting
other hydrides, replacing sulfur with different elements within the
same $Im\overline{3}m$ crystal structure \cite{Papaconstantopoulos2017,Papaconstantopoulos2017a,Fan2016,Heil2015},
which includes isoelectronic counterparts such as Se \cite{Zhang2015b,Flores-Livas2016,Heil2015}.

Although it is commonly accepted that hydrogen is the main contributor
to $T_{c}$ in H$_{3}$S, other hydride-forming elements may have
a dramatic impact on the hydrogen contribution. The purpose of this
work is to present a comprehensive study of the electronic structure
of the hydrides H$_{3}$X (X=As, Se, Br, Sb, Te and I) in the $Im\overline{3}m$
crystal structure and using the GG theory to calculate the Hopfield
parameter $\eta$. To explore the possible superconducting properties
of these materials and compare with the well established H$_{3}$S,
we make estimates of the phonon frequencies from Refs. \cite{Flores-Livas2016,Papaconstantopoulos2015}
and we conclude that a large value of the $\eta$ parameter is the
strongest indication of high T$_{c}$ in these materials as in the
case in H$_{3}$S.

\section{Computational details}

The electronic structure calculations are performed with the all-electron
Linearized Augmented Plane Wave (LAPW) method \cite{Singh1994} specifically
the Wei-Krakauer-Singh code \cite{LAPW1985} developed at the U.S.
Naval Research Laboratory. In the present calculations the Hedin-Lundqvist
form of the local density approximation was used.\cite{HL1971}

To ensure sufficient accuracy for convergence, the total and orbital-projected
densities of electronic states (pDOS) are calculated by the tetrahedron
method with a uniformly distributed k-point grid of 1785 k-points
in the irreducible Brillouin zone.

The key step to estimate $T_{c}$ is the determination of the electron-phonon
coupling $\lambda$, which, as pointed out by McMillan \cite{McMillan1968}
and Hopfield \cite{HOPFIELD1969}, can be written as

\begin{equation}
\lambda_{j}=\frac{\eta_{j}}{M_{j}\left\langle \omega_{j}^{2}\right\rangle }=\frac{N(E_{F})\left\langle I_{j}^{2}\right\rangle }{M_{j}\left\langle \omega_{j}^{2}\right\rangle }\label{eq:lambda}
\end{equation}
where $N(E_{F})$ is the total DOS per spin at the Fermi level $E_{F}$,
$\left\langle I^{2}\right\rangle $ is the electron--ion matrix element,
$\left\langle \omega_{j}^{2}\right\rangle $ is the average phonon
frequency and the index $j$ corresponds to $X$ element and hydrogen.
The Hopfield parameter $\eta_{j}$, which only describes electronic
properties, is calculated using the GG formula based on the scattering
theory. This formula allows us to express the electronic contributions
to the $\lambda_{j}$ in local terms in the following form

\begin{equation}
\eta_{j}=\frac{1}{N(E_{F})}\sum_{l=0}^{2}2(l+1)\sin^{2}(\delta_{l}^{j}-\delta_{l+1}^{j})v_{l}^{j}v_{l+1}^{j}\label{eq:eta}
\end{equation}

where both $\delta_{l}^{j}$ and $v_{l}^{j}=N_{l}^{j}(\epsilon_{F})/N_{l}^{j(1)}$
are orbital l and site j dependent. The phase shifts $\delta_{l}^{j}$
are defined through the following equation:

\begin{equation}
\tan\delta(R_{s},E)=\frac{j_{l}'-j_{l}(kR_{s})L_{l}(R_{s},E)}{n_{l}'-n_{l}(kR_{s})L_{l}(R_{s},E)},\label{eq:phase_tab}
\end{equation}
where $L_{l}=u'_{l}/u_{l}$ is the logarithmic derivative and $j_{l}$
and $n_{l}$ are spherical Bessel and Neumann functions. The free
scatterer DOS $N_{l}^{j(1)}$ is defined as follows: 
\begin{equation}
N_{l}^{j(1)}=(2l+1)\intop_{0}^{R_{s}}\left[u_{l}^{j}(r,E_{F})\right]^{2}r^{2}dr\label{eq:scatter}
\end{equation}
where $u_{l}$ is the radial wave function and the upper limit of
the integral is the muffin-tin radius $R_{s}$.

It should be stressed here that the GG formula Eq. \ref{eq:eta} requires
the use of an all electron potential and therefore it should not be
compatible with pseudopotential methods.

Finally, $T_{c}$ is evaluated using the Allen--Dynes equation \cite{P.B.AllenandR.C.Dynes1975}
as follows:

\begin{equation}
T_{c}=f_{1}f_{2}\frac{\omega_{log}}{1.2}\exp\left[-\frac{1.04(1+\lambda)}{\lambda-\mu^{\star}(1+0.62\lambda)}\right]\label{eq:tc_ad-1}
\end{equation}
In Eq. \ref{eq:tc_ad-1}, $\lambda=\lambda_{X}+3\lambda_{\mathrm{H}}$
where $\lambda_{X}$ represent the acoustic modes of the element X
and $\lambda_{H}$ the optical modes of H. This separation is exact
for these materials and was pointed out long time ago for other hydrides
\cite{Klein1976}. We have set the Coulomb pseudopotential $\mu^{\ast}=0.1$
and $f_{2}=1$. $f_{1}$ is the strong coupling factor given by the
following

\[
f_{1}=\left[1+(\frac{\lambda}{2.46+9.35\mu^{\star}})^{1.5}\right]^{1/3}.
\]

The phonon dispersions and electron-phonon couplings are calculated
using density functional perturbation theory (DFPT) \cite{DFPT} and
the plane-wave pseudopotential method implemented in Quantum-Espresso
package \cite{QE} with ultrasoft pseudopotential \cite{USPP}, kinetic
energy cutoff of 75 Ry and a $24\times24\times24$ k-point and a $6\times6\times6$
q-point mesh. The k-space integrations for DFPT were done with the
tetrahedron method which corresponds to the zero-width condition in
smearing.

\subsection{Results and Analysis}

\subsubsection*{Electronic structure}

\begin{figure}
\centerline{\includegraphics[scale=0.4]{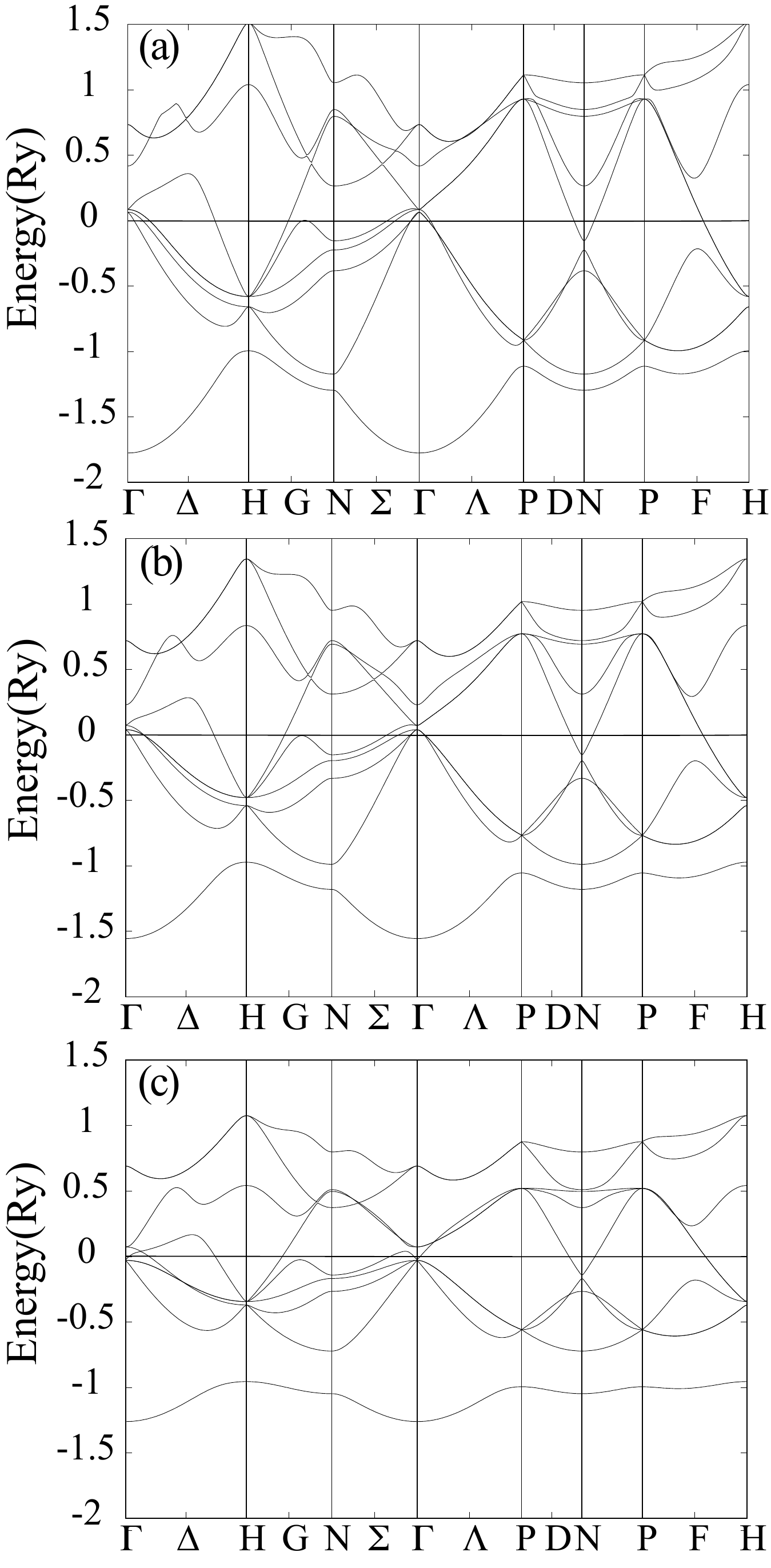}}

\caption{\label{fig:band}Energy bands of H$_{3}$Se under pressure (a) P=2.45
Mbar (b) P=0.76 Mbar and (c) P=0.}
\end{figure}

\begin{figure}
\centerline{\includegraphics[scale=0.4]{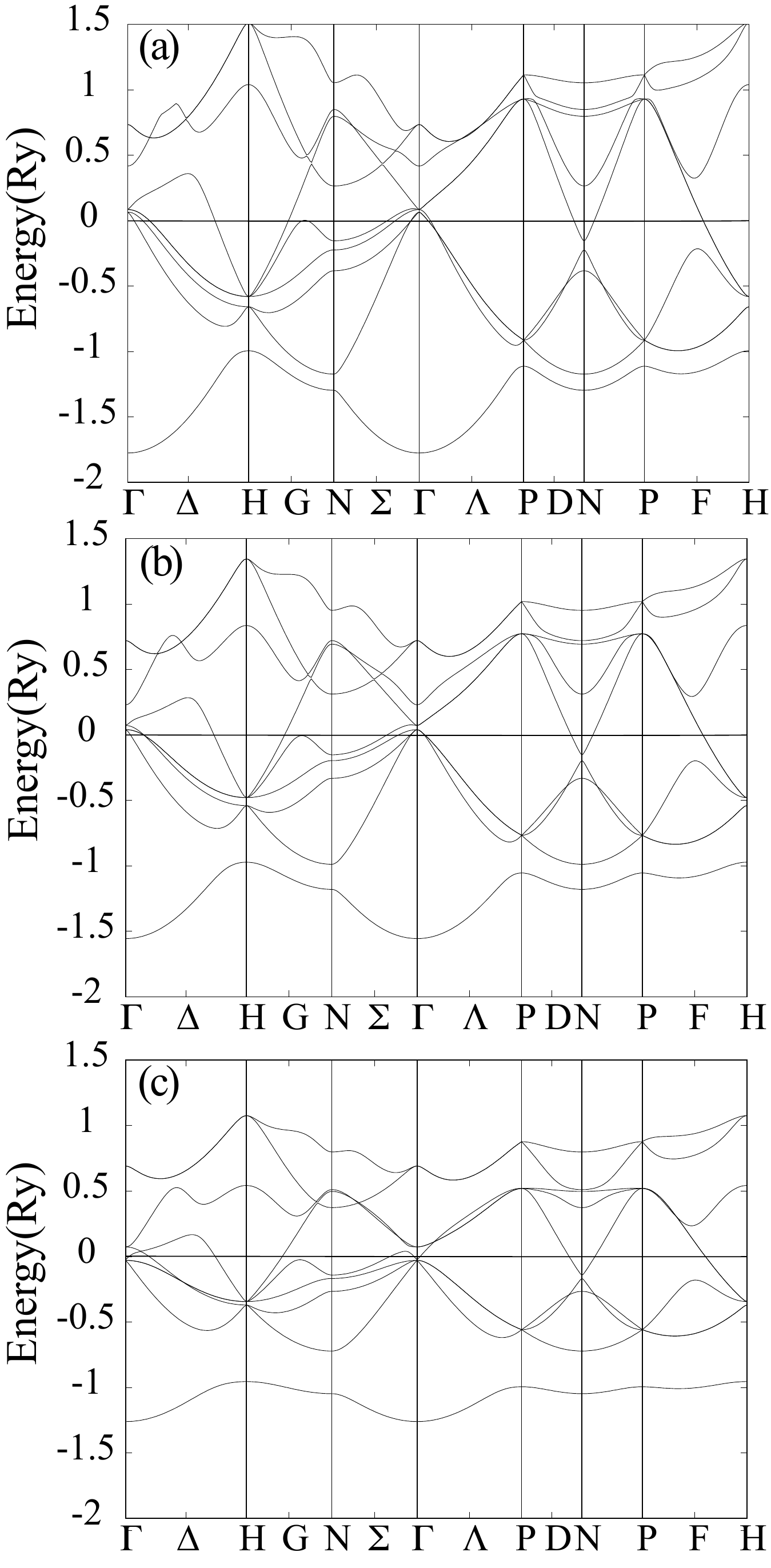}}

\caption{\label{fig:banddos} (a)-(c)H$_{3}$Se Total DOS,(d)-(f) selenium
s-p-d like DOS, (g)-(i) hydrogen s-p like DOS of H$_{3}$Se for a=5.8
a.u., a=6.4 a.u. and 7.2 a.u. that correspond to $P=2.45$ Mbar, $P=0.76$
Mbar and $P=0$, respectively. The Fermi levels are shifted to 0 as
indicated by the vertical line.}
\end{figure}

Fig. \ref{fig:band} shows the energy bands of H$_{3}$Se in the $Im\overline{3}m$
crystal structure \cite{R3M} for three different lattice constants
a=5.8 a.u., a=6.4 a.u. and a=7.2 a.u. that correspond to pressures
P=2.45 Mbar, P=0.76 Mbar and P=0 respectively. Comparing Figs (a),
(b) and (c) we note that for P=0 the lowest band is completely separated
and forms a gap as is also shown in the DOS Fig. \ref{fig:banddos}
(c). This gap gradually closes for P=0.76 Mbar and P=2.45 Mbar. Our
observation of the separated band at $P=0$ is stated for completeness
and not as an explanation of the occurrence of superconductivity at
higher pressure. The overall bandwidth increases significantly with
increasing pressure as expected. However, for the high pressure cases
near the Fermi level $E_{F}$, the ordering and shape of the bands
are not seriously affected. We note that these bands look very similar
to those of the prototype material H$_{3}$S. We have also calculated
the energy bands of the other materials under investigation here i.e.
for X=As, Br, Sb, Te and I. The difference is basically in the position
of $E_{F}$, and therefore we will not present additional band structure
figures but we will come back later to this point on the applicability
of a rigid band behavior in these materials.

Similar information can also be found in Fig \ref{fig:banddos}, which
shows the total DOS (a)-(c), Se-site angular-momentum-decomposed DOS
(d)-(f) and H-site DOS (g)-(i) of H$_{3}$Se where each column corresponds
to $P=2.45$ Mbar, $P=0.76$ Mbar and $P=0$ respectively. Consistent
with Fig.\ref{fig:band}, at high pressures the shape of the DOS is
preserved around $E_{F}$, including the position of $E_{F}$ on a
sharp peak (van Hove singularity) . In the equilibrium case (P=0),
the DOS at the Fermi level is composed of 50\% of p-like Se and 20\%
of s-like H states suggesting a strong sp orbital hybridization. The
d-like Se contribution to the DOS in Figs. \ref{fig:banddos}(d)-(f),
becomes larger as pressure increases. The percentage of d-like states
is doubled from 6\% in equilibrium condition to 13.5\% at P=2.45 Mbar.
A similar pressure-enhanced trend, although much smaller in magnitude,
also appears in the p-like H states as seen in Figs. \ref{fig:banddos}(g)-(i).

\begin{figure}
\centerline{\includegraphics[scale=0.55]{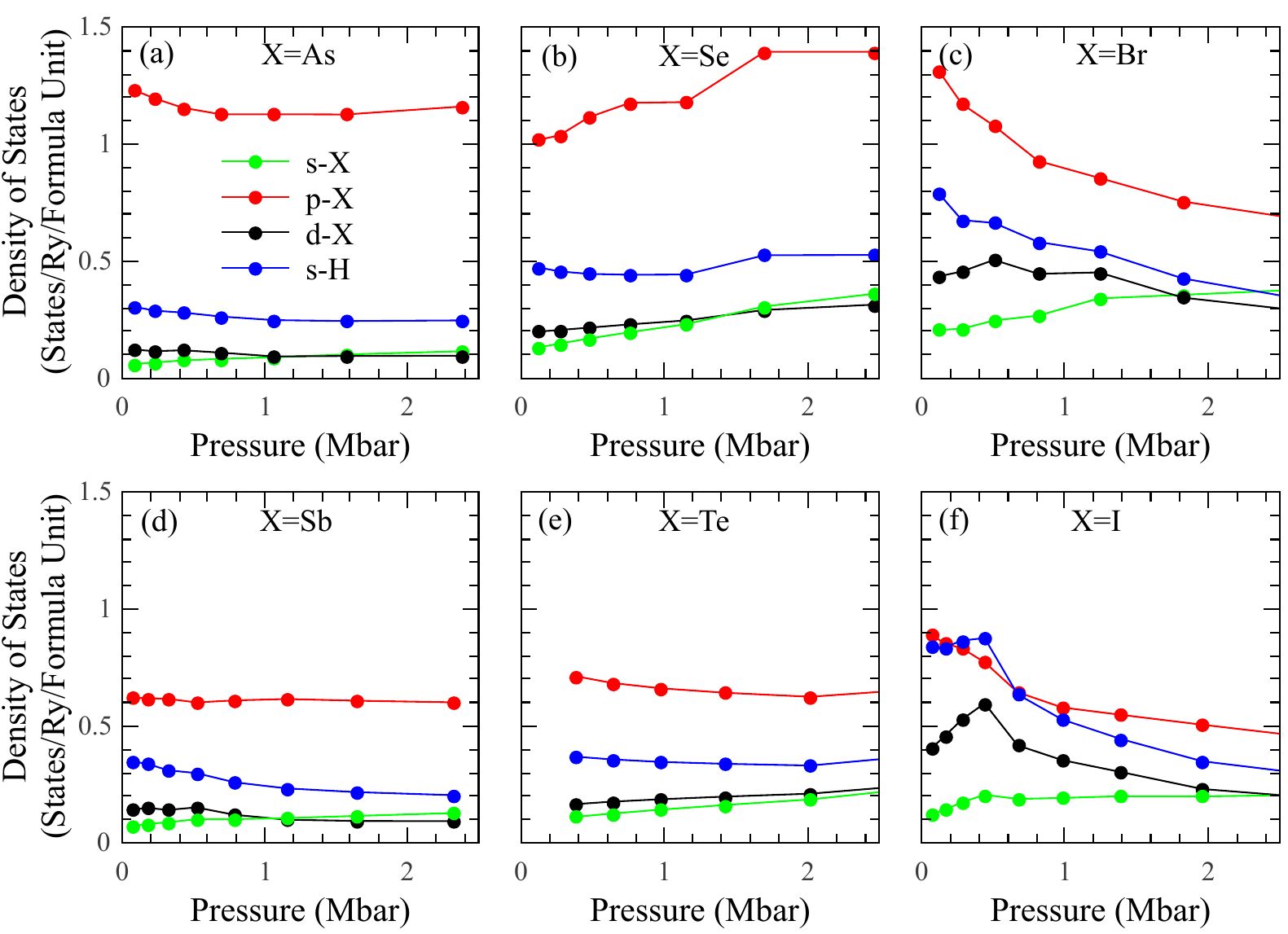}}

\caption{\label{fig:pdos-p} Projected DOS of H$_{3}$X at Fermi level vs.
Pressure where (a)-(f) each corresponds to X=As, Se, Br, Sb, Te and
I respectively. In the second row smaller DOS values are seen.}
\end{figure}

Fig. \ref{fig:pdos-p} shows the $\ell$-components of the DOS at
the Fermi level vs pressure for H\textbf{$_{3}$}X (X=As, Se and Br
in the top row and Sb, Te and I in the bottom row). A tabulation of
these results is found in Table I of the Appendix. All six materials
show the p-like X component to be the dominant one with the $d_{X}$
and $s_{H}$ to have 2-3 times smaller values. However, the $d_{X}$
which is the smallest component in the equilibrium condition ($P=0$),
increases monotonically with the fastest rate in all cases. The important
finding here is that the $p_{H}$ becomes significant at the high
pressures where superconductivity occurs. We note that we present
these results at low pressures to show the trends of the projected
DOS within the $Im\overline{3}m$ structure. However for these materials
this structure is not stable at low pressures, as discussed in Ref.
\cite{Einaga2016} the stable crystal structure is in the rhombohedral
R3m space group \cite{R3M} .

\subsubsection*{Hopfield parameter $\eta$}

\begin{figure}
\centerline{\includegraphics[scale=0.5]{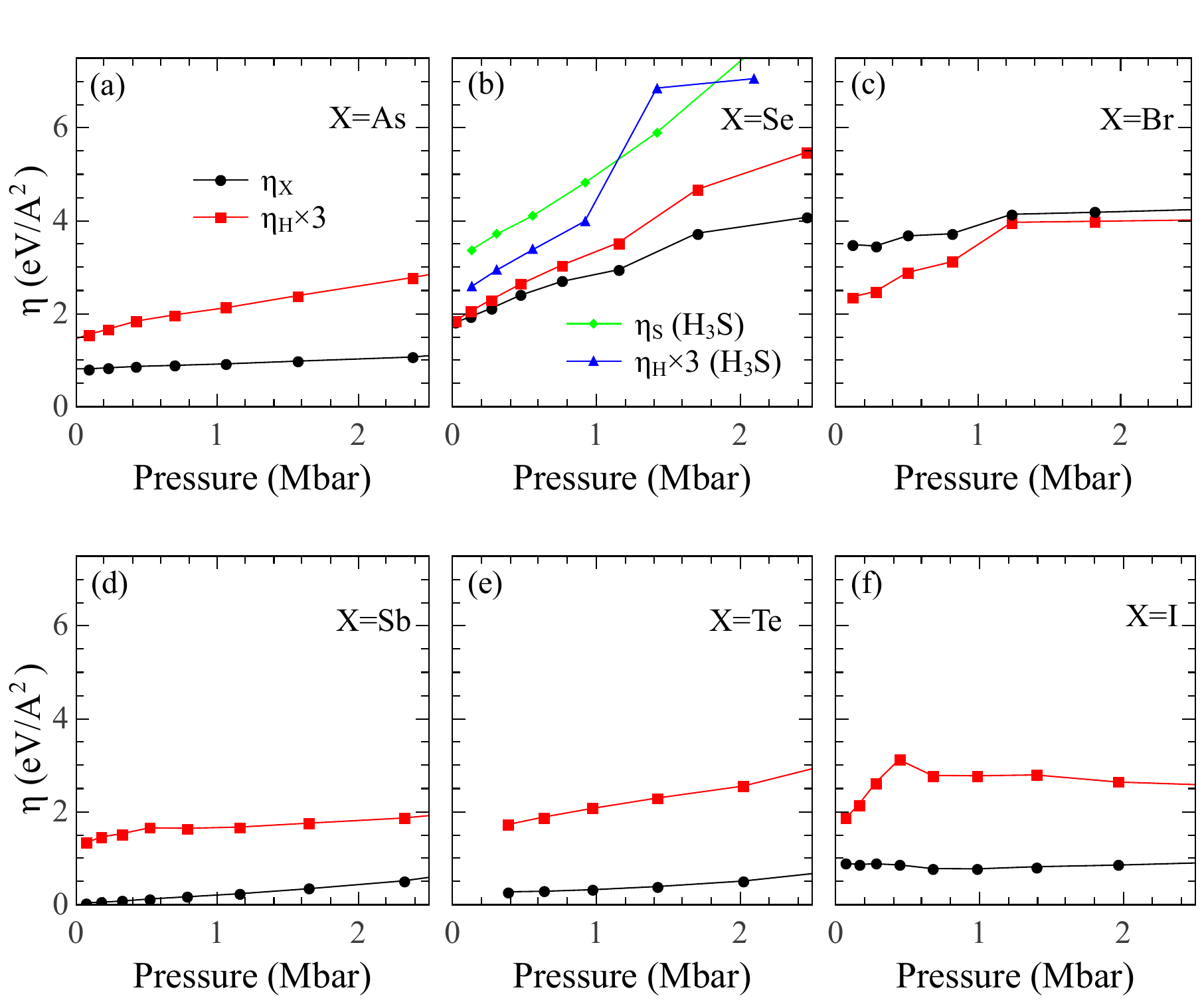}}

\caption{\label{fig:eta}Hopfield parameters $\eta$ vs pressure for H$_{3}$X.
(a)-(f) correspond to X=As, Se, Br, Sb, Te and I respectively. $\eta_{H}$
has been multiplied by three. In (b) we also show $\eta$ for H$_{3}$S
for comparison.}
\end{figure}

\begin{figure}
\centerline{\includegraphics[scale=0.58]{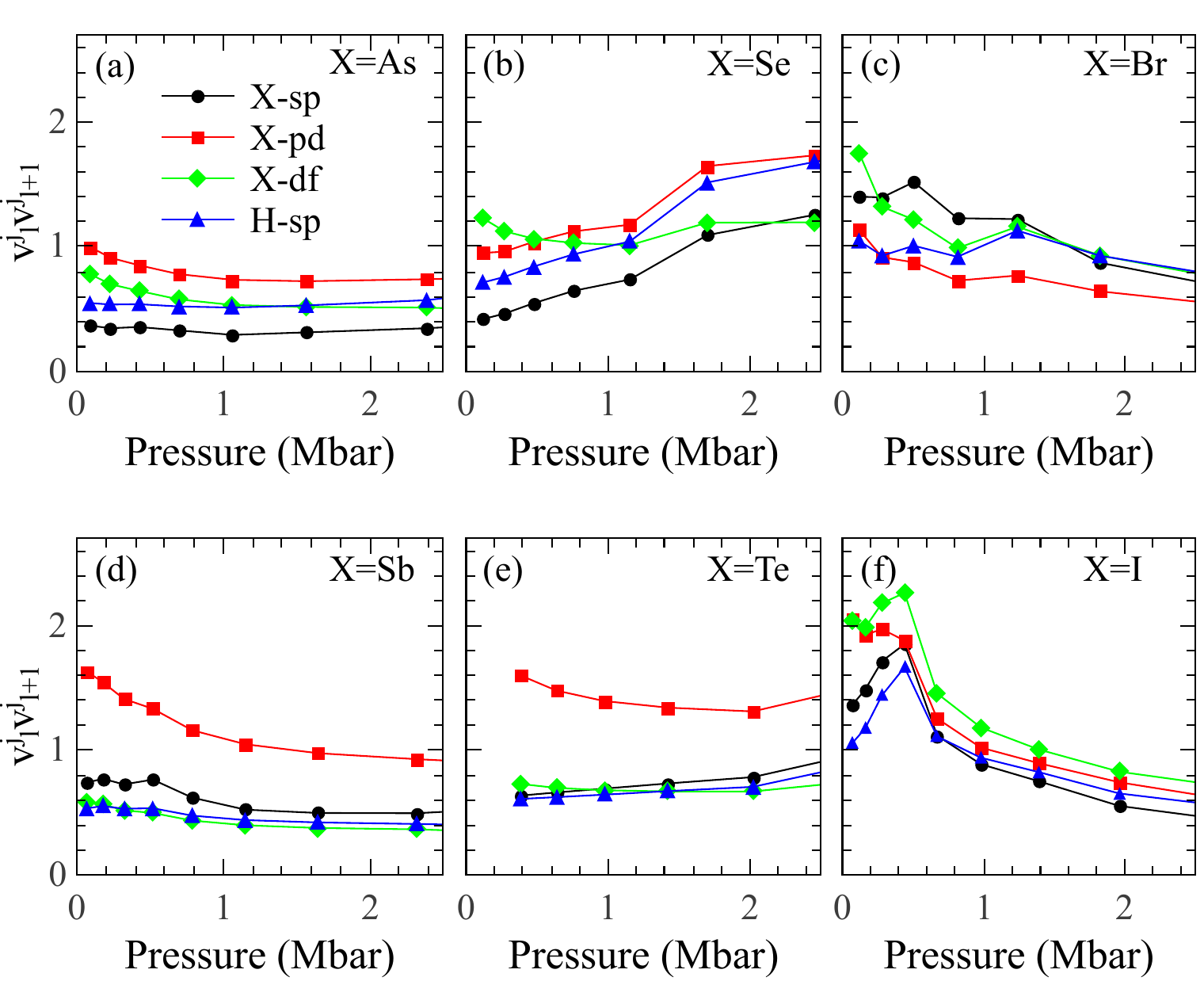}}

\caption{\label{fig:pdos-p-1}partial-DOS product terms in Eq. \ref{eq:eta}
vs. Pressure for H$_{3}$X at Fermi level. (a)-(f) each corresponds
to X=As, Se, Br, Sb, Te and I respectively. Similar features as in
Fig. \ref{fig:pdos-p} for the DOS are captured.}
\end{figure}

\begin{figure}
\centerline{\includegraphics[scale=0.58]{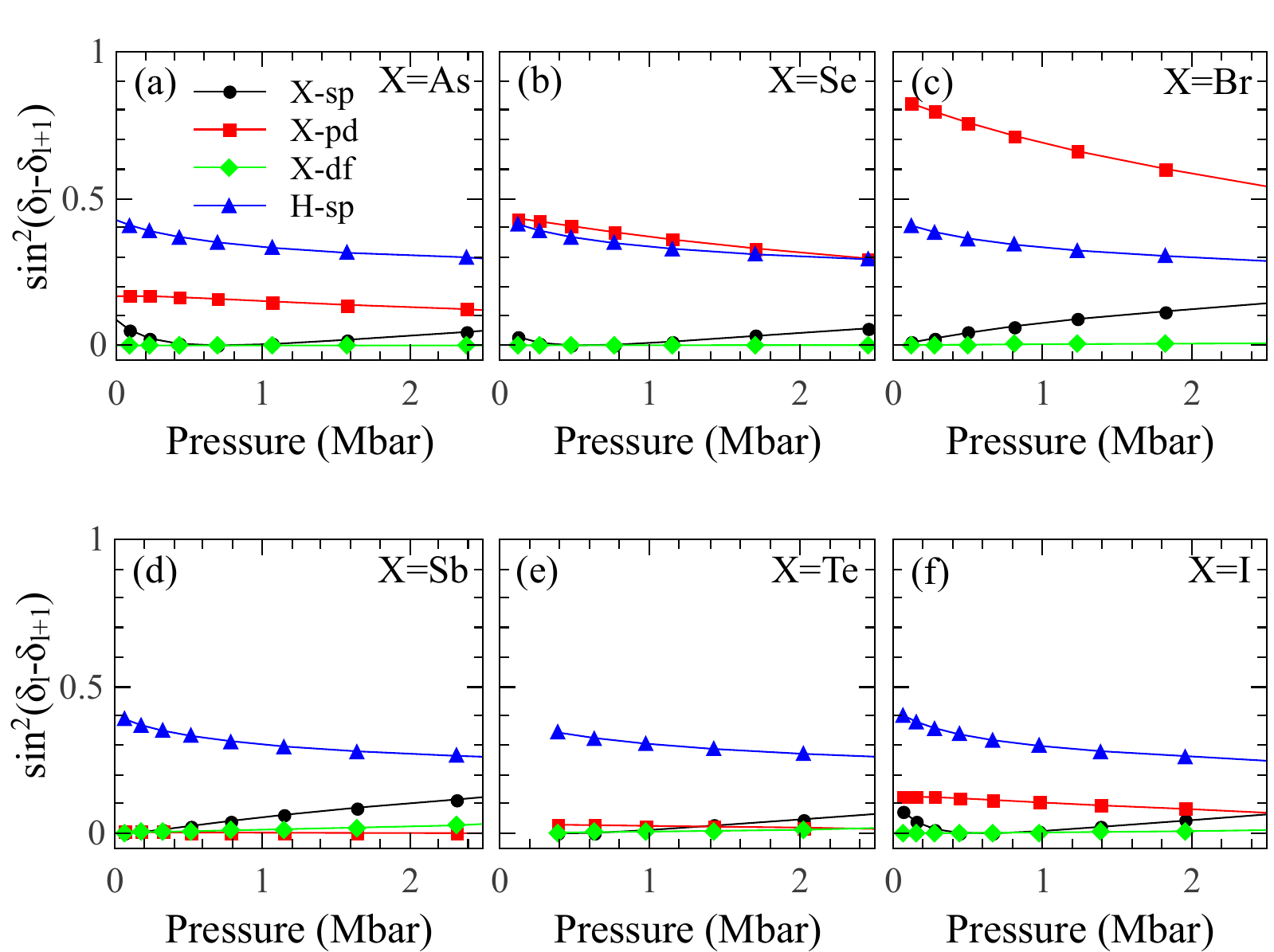}}

\caption{\label{fig:pdos-p-1-1} Phase shift related factor $\sin^{2}(\delta_{l}^{j}-\delta_{l+1}^{j})$
vs. Pressure for H$_{3}$X at Fermi level. (a)-(f) each corresponds
to X=As, Se, Br, Sb, Te and I respectively.}
\end{figure}

We have calculated the Hopfield parameter $\eta_{j}$ using Eq. \ref{eq:eta}
in Section II. In this formulation the index j indicates that we obtain
separate $\eta_{j}$ for hydrogen and the element X. This also results
in having two separate electron-phonon coupling constants $\lambda_{j}$
which give total $\lambda=\lambda_{X}+3\lambda_{\mathrm{H}}$. This
approach differs from the approach of other authors, who directly
compute the total $\lambda$. Our approach has the advantage of studying
the electronic contribution to $\lambda$ from each element separately
and being able to pin down which aspects of the band structure affect
superconductivity as is described below. The summation Eq. \ref{eq:eta},
in a cubic approximation, has three terms which we identify as the
sp (for l=0), pd (for l=1) and df (for l=2) channels. For the hydrogen
component $\eta_{H}$ the dominant term comes from the sp channel.
For the X component, $\eta_{X}$ of the H$_{3}$X compounds the dominant
term comes from the pd channel. It should also be noted that each
term of the sum consists of the product $\sin^{2}(\delta_{l}^{j}-\delta_{l+1}^{j})v_{l}^{j}v_{(l+1)}^{j}$.
The $v_{l}^{j}v_{(l+1)}^{j}$ term of the product is usually larger
but the $\sin^{2}(\delta_{l}^{j}-\delta_{l+1}^{j})$ is not negligible.
In Fig. \ref{fig:eta} $\eta_{j}$ is plotted versus pressure for
the six materials we have studied. The values of $\eta_{H}$ have
been multiplied by three because of the three crystallographic sites
of hydrogen. From the six compounds, it is found that H$_{3}$Se has
the largest values of $\eta$ but significantly smaller than those
of the prototype material H$_{3}$S as shown in Fig. \ref{fig:eta}
(b). The others have lower values of $\eta$ especially those of the
second row. We now proceed to analyze the relative importance of the
two terms of the product shown in Fig. \ref{fig:pdos-p-1} and \ref{fig:pdos-p-1-1}.

\begin{figure}
\centerline{\includegraphics[scale=0.53]{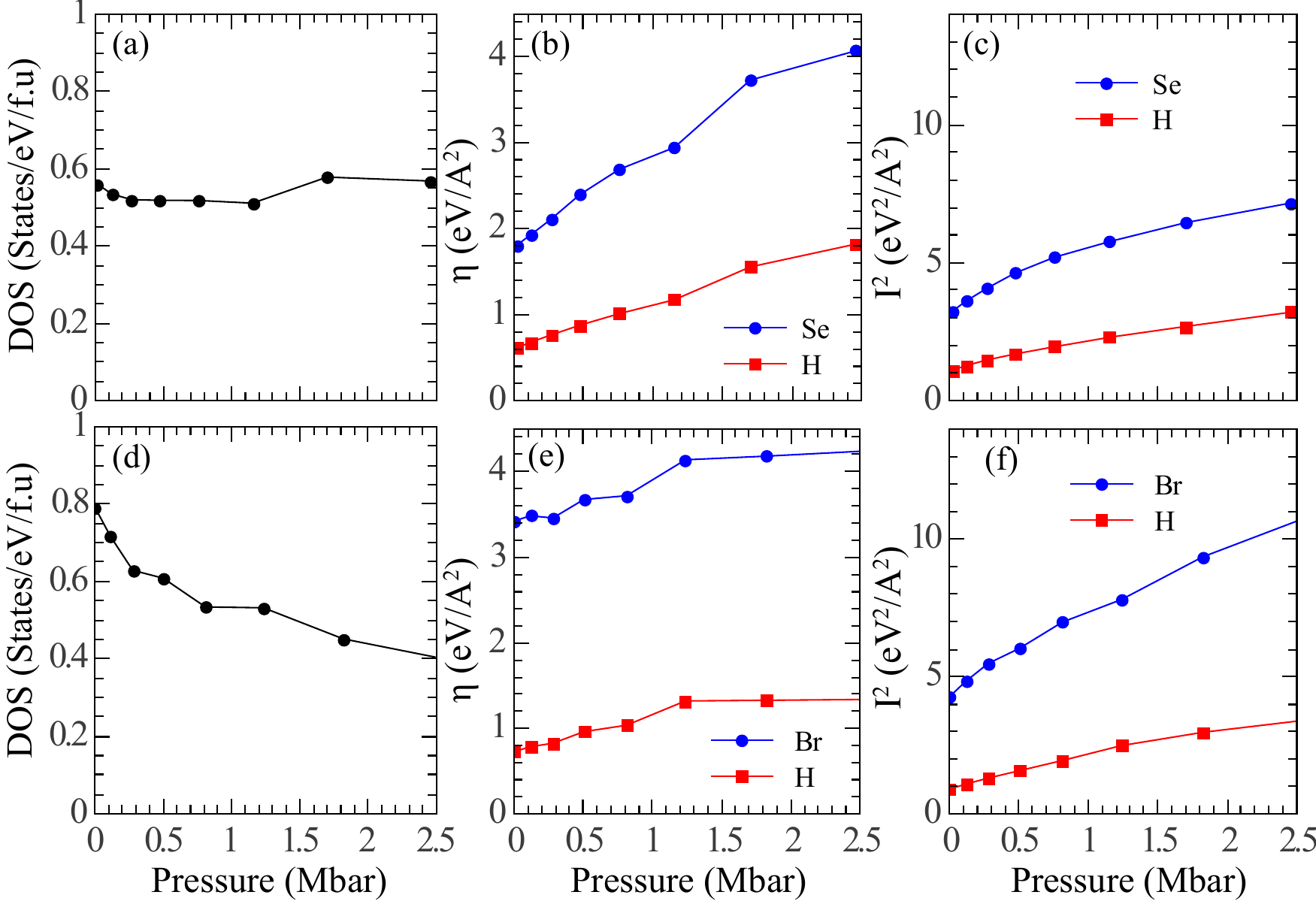}}

\caption{\label{fig:I2} (a)-(c) corresponds to $N(E_{F})$, $\eta_{j}$ and
$\left\langle I_{j}^{2}\right\rangle $ of H$_{3}$Se and (d)-(f)
corresponds to those of H$_{3}$Br respectively.}
\end{figure}

The term $v_{l}^{j}v_{(l+1)}^{j}$ in Eq. \ref{eq:eta} plotted against
pressure is shown in Fig. \ref{fig:pdos-p-1}. One can see it retains
the general trends and certain features such as few jumps and flatness
at various pressure by comparing Fig. \ref{fig:pdos-p-1} to Fig.
\ref{fig:pdos-p}. Although the difference between X=Se and Te is
more pronounced than others which can be attributed to the relative
location of Fermi level to the peak in DOS, the rest are roughly in
the same magnitude.

The second factor, $\sin^{2}(\delta_{l}^{j}-\delta_{l+1}^{j})$, on
the other hand, which describes the effect of phase shift as shown
in Fig. \ref{fig:pdos-p-1-1}, amplifies the difference and its influence
can be summarized by the following trends. For H-site, while the $\sin^{2}(\delta_{l}^{j}-\delta_{l+1}^{j})$
factor in the upper row is generally larger than that in the bottom
row, $\sin^{2}(\delta_{0}^{H}-\delta_{1}^{H})$ within the same row
are nearly identical. For X elements the most significant contribution
to $\sin^{2}(\delta_{1}^{X}-\delta_{2}^{X})$ term corresponds to
the pd channel (with square symbols) in Fig. \ref{fig:pdos-p-1-1},
going from H$_{3}$As to H$_{3}$Br we note a substantial increase
with pressure, while being vanishingly small in all three materials
in the bottom row. The phase shifts depend on the logarithmic derivatives
$L_{l}$ in Eq. \ref{eq:phase_tab} which in turn depend on the crystal
potential that strongly varies from one material to another. Finally,
from Eq. \ref{eq:lambda} we see that the Hopfield parameter is defined
as the product of the total DOS at $E_{F}$, $N(E_{F})$, and the
electron-ion matrix elements $\left\langle I_{j}^{2}\right\rangle $
for each of the two components X and H.

As an example we plot the quantities $N(E_{F})$, $\eta_{X}$ and
$\eta_{H}$ for X=Br and Se in Fig. \ref{fig:I2}. we note that although
$\eta$'s generally increase with increasing pressure, the underlying
reasons are a bit different. In the case of H$_{3}$Se, $N(E_{F})$
is slowly varying with pressure while $\eta$ and $\left\langle I_{j}^{2}\right\rangle $
have a rapid increase with pressure. In the case of H$_{3}$Br, on
the other hand, the $N(E_{F})$ decreases rapidly and competes with
increasing $\left\langle I_{j}^{2}\right\rangle $. However, the important
message is that the increase of $\left\langle I_{j}^{2}\right\rangle $
dominates over the $N(E_{F})$ with the resulting $\eta$ to always
increase with pressure.

\subsubsection*{Phonon frequencies and $T_{c}$}

\begin{table*}
\fontsize{9}{12}\selectfont

\begin{tabular}{>{\centering}p{1.1cm}|>{\centering}p{1.1cm}|>{\centering}p{1.1cm}>{\centering}p{1.1cm}>{\centering}p{1.5cm}>{\centering}p{1.5cm}|>{\centering}p{1.1cm}>{\centering}p{1.1cm}>{\centering}p{1.1cm}|>{\centering}p{1.1cm}>{\centering}p{1.1cm}>{\centering}p{1.1cm}|>{\centering}p{1.1cm}}
 & $N(E_{F})$  & $\ensuremath{\eta_{X}}$  & $\ensuremath{\eta_{\mathrm{H}}}$  & $M_{X}\left\langle \omega_{\mathrm{X}}^{2}\right\rangle $  & $M_{\mathrm{H}}\left\langle \omega_{\mathrm{H}}^{2}\right\rangle $  & $\omega_{log}$  & $\sqrt{\left\langle \omega_{\mathrm{X}}^{2}\right\rangle }$  & $\sqrt{\left\langle \omega_{\mathrm{H}}^{2}\right\rangle }$  & $\lambda_{\mathrm{X}}$  & $3\lambda_{\mathrm{H}}$  & $\lambda$  & $T_{c}$\tabularnewline
 & $\mathrm{\frac{states}{Ry}}$ & \multicolumn{4}{c|}{(eV$/\mathring{A}^{2}$)} & \multicolumn{3}{c}{(K)} &  &  &  & (K)\tabularnewline
\hline 
H$_{3}$S  & 8.78 & 7.68  & 2.29  & 17.08  & 4.11 & 1348  & 546  & 1514  & 0.46  & 1.68  & 2.12 & 231\tabularnewline
H$_{3}$As  &  & 1.03  & 0.87  &  &  &  &  &  &  &  &  & \tabularnewline
H$_{3}$Se  & 6.70  & 3.90  & 1.68  & 17.62  & 5.60  & 1357  & 353  & 1762  & 0.22  & 0.91  & 1.13  & 118\tabularnewline
H$_{3}$Br  & 5.85  & 4.19  & 1.33  & 13.84  & 4.22  & 986  & 311  & 1536  & 0.30  & 0.96  & 1.25  & 98\tabularnewline
H$_{3}$Sb  &  & 0.44  & 0.60  &  &  &  &  &  &  &  &  & \tabularnewline
H$_{3}$Te  &  & 0.48  & 0.85  &  &  &  &  &  &  &  &  & \tabularnewline
H$_{3}$I  &  & 0.84  & 0.87  &  &  &  &  &  &  &  &  & \tabularnewline
\end{tabular}

\caption{\label{tab:lambda} N($E_{F}$) and $\eta_{j}$ calculated using LAPW
and GG theory for H$_{3}$X (X=As, Se, Br, Sb, Te and I) and $\lambda_{j}$,
$\omega_{log}$ and $\sqrt{\left\langle \omega_{\mathrm{j}}^{2}\right\rangle }$
calculated using QE for X=S, Se and Br.}
\end{table*}

We retain the two-component approach separating the acoustic from
the optic modes in these materials as justified by the small mass
of hydrogen \cite{Klein1976} and also verified by the lattice dynamics
calculations of other groups \cite{Flores-Livas2016,Duan2014}. More
importantly, we follow McMillan's classic equation which separates
the electron-phonon coupling constant $\lambda_{j}$ into a numerator
$\eta_{j}$ which represents the electronic contribution and a force
constant $M_{j}\left\langle \omega_{\mathrm{j}}^{2}\right\rangle $
in the denominator representing the phononic contribution. This separation
is advantageous because it provides insights into understanding the
reason superconductivity occurs in these materials.

The Hopfield-McMillan parameter $\eta_{j}$ we analyzed in the previous
section which we calculated directly by the Gaspari-Gyorffy theory,
and identified the importance of the different terms in the GG formula.
In order to calculate the force constants we recast McMillan Eq. \ref{eq:lambda}
into the following form

\begin{equation}
M_{j}\left\langle \omega_{j}^{2}\right\rangle =\frac{\eta_{j}}{\lambda_{j}}.\label{eq:w2}
\end{equation}
where $\lambda_{j}$ is calculated from independent Quantum-Espresso
for H$_{3}$S, H$_{3}$As, H$_{3}$Se and H$_{3}$Br. The QE calculations,
in addition, give $\omega_{log}$ needed for $T_{c}$. The results
from both QE and the GG theory are summarized in Table \ref{tab:lambda}.
The results of H$_{3}$As from QE show negative phonon frequencies
rendering this material as unstable.

Our QE results for both H$_{3}$Se and H$_{3}$S give values for $\omega_{log}$
and $\lambda$ very close to previous works Refs. \cite{Flores-Livas2016,Duan2014}.
A small difference between our estimated $T_{c}$ and Ref.\cite{Duan2014}
is partly due to a different value of $\mu^{\varhexstar}=0.1$ used
in our calculations and because we also used a more general Allen-Dynes
expression Eq. \ref{eq:tc_ad-1} including the mutiplier $f_{1}$,
which is more suitable for large $\lambda$ \cite{GorKov2018}, rather
than McMillan-Dynes expression \cite{McMillan1968,DYNES1972615} implemented
in QE by default.

In Table \ref{tab:lambda} we show $N(E_{F})$ calculated by the LAPW
method, $\eta_{j}$ by the GG theory, $\lambda_{j}$, $\omega_{log}$
and $\omega_{j}$ by the QE code.

This Table \ref{tab:lambda} shows that hydrogen is the main contributor
to the $T_{c}$ in all three calculated materials as it makes up more
than $75$\% of the total $\lambda$ while providing higher frequency
vibrational modes.

The optical phonon frequency $\left\langle \omega_{H}^{2}\right\rangle ^{1/2}$
is generally insensitive to the X element that hydrogen forms hydrides
with as long as the $M_{x}$ is large enough to ensure the phonon
mode separation. The difference in $\left\langle \omega_{H}^{2}\right\rangle ^{1/2}$
between H$_{3}$Se and H$_{3}$S is less than $10\%$. The lower frequency
acoustic branch $\left\langle \omega_{X}^{2}\right\rangle ^{1/2}$
on the other hand, varies more significantly, as X represents a different
element in a different hydride, depending on the bonding and the atomic
mass of the X element. As expected $\omega_{log}$ could also have
noticeable change as it accounts for the collective behavior of all
elements in the material. However the influence of the phonon frequencies
on $T_{c}$ is more limited due to the fact that $\omega_{log}$ and
$\left\langle \omega_{j}^{2}\right\rangle ^{1/2}$ enter different
parts of Eq. \ref{eq:tc_ad-1} and have opposite effect on $T_{c}$
as they are correlated.

For instance, while having slightly smaller $\eta$'s in H$_{3}$Br
($\eta_{Se}=4.19$ and $\eta_{H}=1.33$ eV$/\mathring{A}^{2}$) than
H$_{3}$Se, the $\lambda$ of H$_{3}$Br is about 10\% larger than
that of H$_{3}$Se at the $P=2$ Mbar. This is mainly because H$_{3}$Br
has smaller $\omega_{j}$'s in the denominator of Eq. \ref{eq:lambda}.
The decrease of $\omega_{j}$ in H$_{3}$Br which leads to the increase
of $\lambda$, will also reflect on the prefactor $\omega_{log}$
in Eq. \ref{eq:tc_ad-1} which lowers the $T_{c}$. As a result, with
similar $\eta$ values in both H$_{3}$Se and H$_{3}$Br, having nearly
32\% larger $\omega_{log}$ in H$_{3}$Se only gives 18\% difference
in its $T_{c}$.

The importance of $\omega_{log}$ in high $T_{c}$ superconductivity
is sometimes overemphasized. H$_{3}$Se and H$_{3}$S have very similar
$\omega_{log}$, however the $T_{c}$ of H$_{3}$S is higher by more
than 100 K. On the other hand, as discussed earlier, H$_{3}$Se and
H$_{3}$Br having similar $\eta$ values leads to vary close $T_{c}$
despite the significant $\omega_{log}$ difference.

The reason becomes clear by comparing the $\eta_{j}$ of H$_{3}$X
(X=Se, Br and S) in Fig. \ref{eq:eta}. It is obvious that only H$_{3}$S
has distinctively large $\eta_{j}$ and therefore allows the system
to have both large $\omega_{log}$ and $\lambda$.

Although from a numerical standpoint, a high T$_{c}$ is rather the
optimal condition of the interplay between $\lambda_{j}$ and $\omega_{log}$
as in the case of H$_{3}$S. The parameter $\eta$, which depends
solely on the electronic structure, is the only factor that can be
optimized independently

The value of the Hopfield parameters $\eta_{j}$ can in principle
change by a lot more because they depend on the Fermi level values
of the angular momentum decomposed electronic densities of states
and how close to a van Hove singularity the Fermi level is. So our
finding is that the Hopfield parameter is a quantity more sensitive
from material to material than the average phonon frequency. Therefore,
our conclusion is that within the $Im\overline{3}m$ crystal structure
a significant increase of the Hopfield parameter from its value in
H$_{3}$S is needed in order to raise $T_{c}$ in the direction of
room temperature. We believe that values of $\eta_{j}$ larger than
10 eV/$\text{Å}^{2}$ are needed for higher $T_{c}$ than in H$_{3}$S.

Our approach based on the GG theory has the advantage of keeping the
decoupling of the electronic component of $\lambda$ from the phonon
component. In addition, separating the contribution of the element
X from that of hydrogen identifies the distinct contributions of acoustic
and optic modes. So those two decouplings i.e. separation of electronic
from phononic contributions and separation of the two elements X and
H offer more insights in the understanding of superconductivity in
these materials.

Regarding the three materials in the second row we found small values
of $\eta_{j}$ and we conclude that given their larger masses it is
unlikely that their force constants would be be small enough to raise
$\lambda$ to the desired value for high $T_{c}$. Therefore we did
not carry out phonon spectra calculations for those.

\subsubsection*{Rigid-band model}

\begin{figure}
\centerline{\includegraphics[scale=0.5]{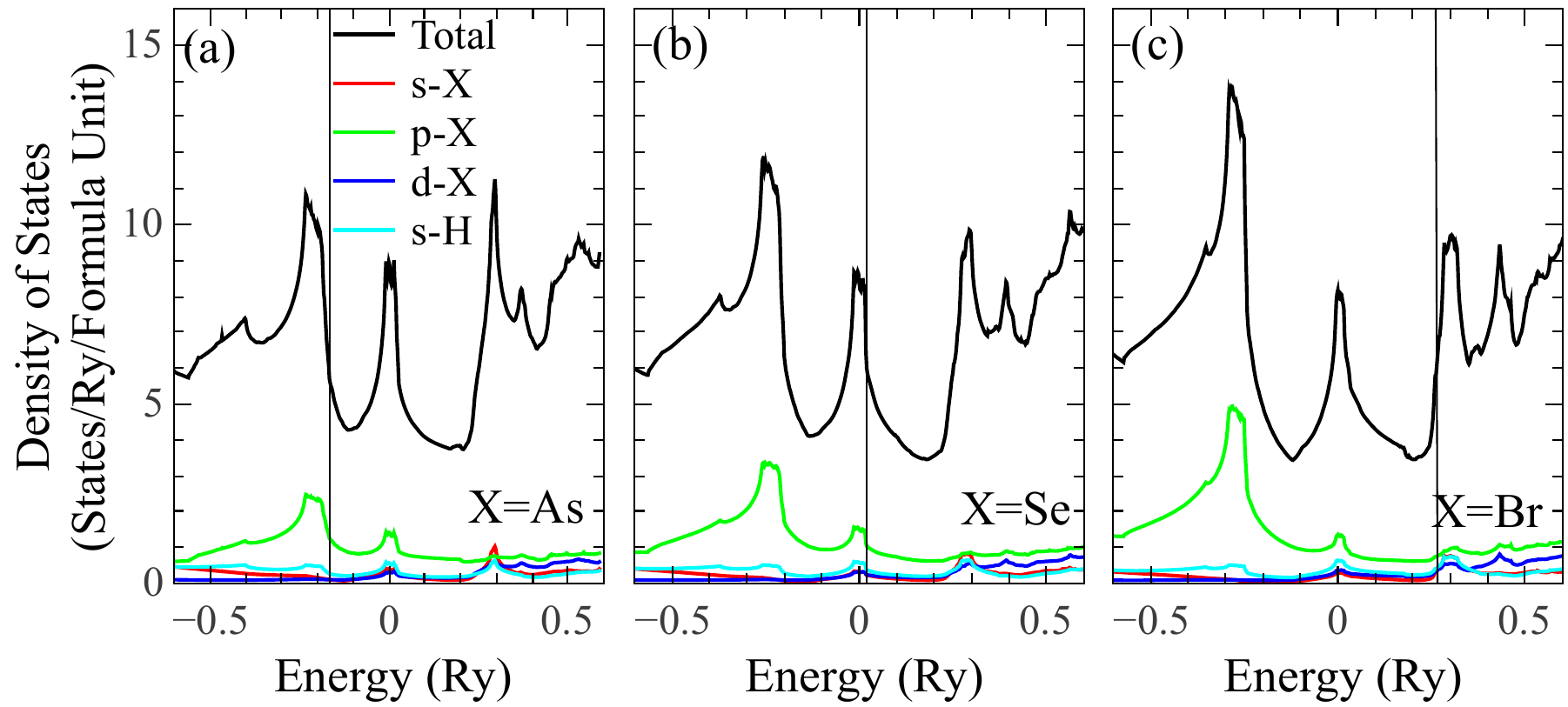}}

\caption{\label{fig:rigidband}DOS for H$_{3}$X (X=As, Se and Br). The DOS
within $\pm$ 0.3 (Ry) around $E_{F}$ is preserved well across adjacent
X elements belonging to the same period.}
\end{figure}

\begin{figure}
\centerline{\includegraphics[scale=0.5]{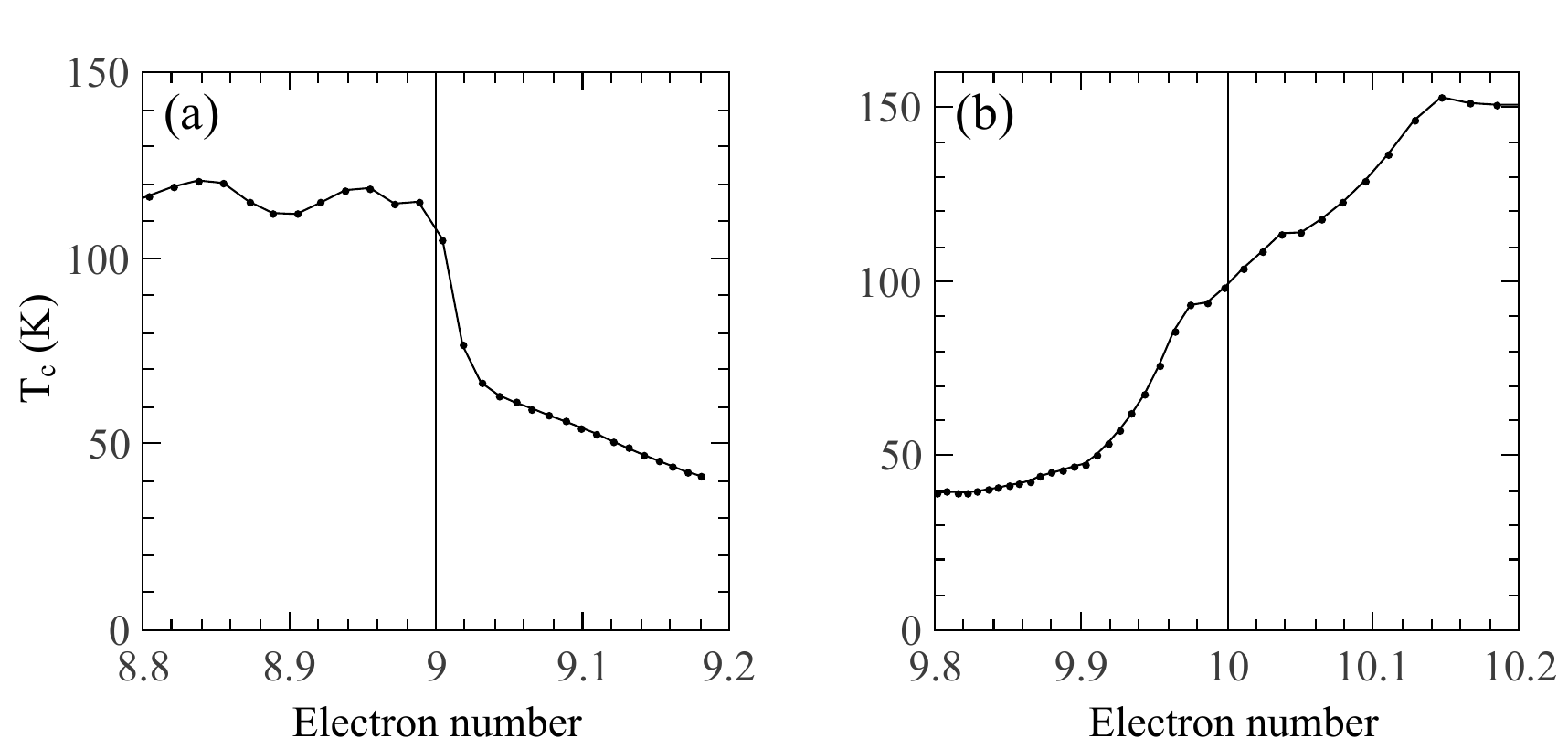}}

\caption{\label{fig:tc_e}Estimated T$_{c}$ in the case of alloying using
rigid band model for (a) H$_{3}$Se and (b) H$_{3}$Br.}
\end{figure}

Fig. \ref{fig:rigidband} shows the total DOS near $E_{F}$ and partial
DOS for all three materials in the first row (X=As, Se and Br) at
Pressure around 2 Mbar. By comparing the three panels, one can see
that the shape of DOS around $E_{F}$ is well preserved. The relevant
electronic properties can thus be explained with a rigid-band model
where the change in electronic states corresponds directly to Fermi
level shift due to a small amount of alloying and indirectly affect
$T_{c}$ through the parameter $\eta_{j}$.

To better understand how the electronic properties could potentially
affect $\lambda$ and $T_{c}$, we also consider $T_{c}$ versus electron
number within a rigid band model. This is shown in Fig. \ref{fig:tc_e}.
Each material has its Fermi level near a peak in the density of states,
even without alloying. By shifting the Fermi level toward the peak,
the $\lambda$ and $T_{c}$ can be enhanced dramatically as in the
case of H$_{3}$Br. However for H$_{3}$Se the enhancement is limited
since $E_{F}$ is very close to the van Hove singularity.

This gives an interpretation on the role of the element X from a different
perspective and provides certain guidelines for optimizing hydrides
to achieving high $T_{c}$. Of special interest is the case of H$_{3}$Br,
where a small amount of additional electrons can enhance the $T_{c}$
to above $150$ K. This may be accomplished by increasing the hydrogen
content in H$_{3}$Br. We confirmed this rigid band prediction by
performing a virtual crystal calculation by increasing the hydrogen
amount by 0.15 electrons.

It should be mentioned here that the idea of substitution of the element
X has been applied in the H-S-Se system by Liu et all \cite{PhysRevB.98.174101}
who discovered three dynamically stable structures which keep the
main features of the cubic $Im\bar{3}m$ structure. Along the same
lines, Amsler \cite{PhysRevB.99.060102} using cluster expansion method
reached the same conclusions and that $T_{c}$ cannot be raised beyond
its value in H$_{3}$S because of the Fermi level moving away from
the van Hove singularity.

\subsubsection*{}

\section{Conclusions}

We have calculated the parameters determining superconductivity in
the group of hydrides H$_{3}$X (X=As, Se, Sb, Te and I). Our approach
combines LAPW electronic-structure calculations, which yield the total
and angular momentum decomposed DOS in the $Im\overline{3}m$ crystal
structure, with the Qauntum-ESPRESSO code from which the phonon spectra
are calculated

We also use the Gaspari--Gyorffy theory to evaluate the McMillan-Hopfield
parameter $\eta_{j}$ and obtain additional insight on the mechanism
of superconductivity in these materials. We conclude that the elements
X play the role of stabilizer of these compounds to keep H metallized
under pressure. The highest superconducting transition temperature
$112$ K is found for H$_{3}$Se, which is isoelectronic to the well-established
H$_{3}$S.

For H$_{3}$Br we calculate a little smaller $T_{c}=98$ K, but using
a rigid band model and a VCA calculation we predict $T_{c}\sim150$
K with electron doping

We find that the road to high $T_{c}$ depends mainly on high values
of the Hopfield parameters on the hydrogen sites. Our analysis shows
that the variation of the parameters $\eta_{H}$ or of the matrix
element $\left\langle I_{H}^{2}\right\rangle $ is more important
than the variation of the phonon frequency.

This is consistent with the view of Pickett and Eremets \cite{phystoday}
who argue that ``obtaining and understanding, and thereby control,
of $\left\langle I^{2}\right\rangle $ is one of the most important
remaining questions in researcher's quest to further increase $T_{c}$
or to reduce the necessary pressure''

Finally, in Refs. \cite{Fan2016,Ge2016} the possibility of achieving
higher $T_{c}$ in H$_{3}$S by adding small amount of phosphorus
is explored with conflicting results. Also, Heil and Boeri \cite{Heil2015}
suggest raising T$_{c}$ in H$_{3}$S by replacing sulfur with oxygen.
In our present study we did not find ways to raise $\left\langle I^{2}\right\rangle $
or $\eta$ resulting in a higher $T_{c}$ in the group of compounds
H$_{3}$X that we used. Ref. \cite{Papaconstantopoulos2017a} reports
a very large $\eta_{F}=17.5$ eV/$\text{Å}^{2}$ for H$_{3}$F, but
the stability of this material is in question. In a preliminary calculation
for H$_{3}$O we have found a very large value of $\eta_{0}=18.4$
eV/$\text{Å}^{2}$ without exploring the stability of the $Im\overline{3}m$
structure.

\section*{ACKNOWLEDGMENTS}
\begin{acknowledgments}
We thank B. M. Klein and W. E. Pickett for useful discussions, and
M. Kawamura for help with DFTP calculations with Quntum-ESPRESSO.
This work was partially supported by the U.S. Department of Energy
grant DE-SC0014337 and by the Alliance for Sustainable Energy. M.
J. Mehl is supported by the Kinnear Foundation and the U.S. Office
of Naval Research via Duke University subaward 313-0710.
\end{acknowledgments}

\section*{Appendix A}

\begin{table*}[t]
\centering

\includegraphics{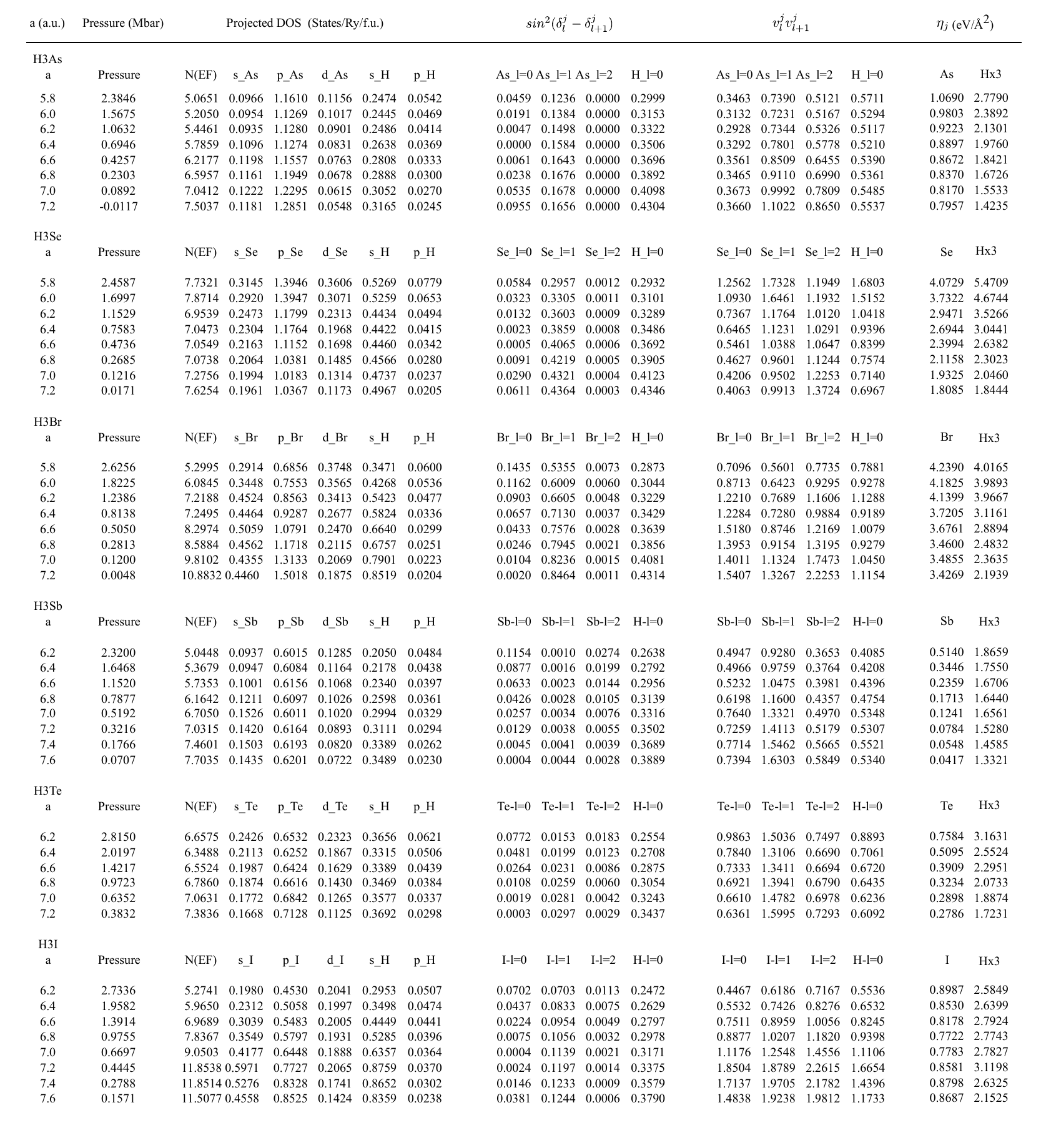}

\caption{\label{tab:appd}Total and projected DOS, Hopfield parameter $\eta$
and the two terms of the products in Eq. \ref{eq:eta}}
\end{table*}

 \bibliographystyle{unsrt}
\bibliography{Tc_lib}
  
\end{document}